# Anatomical MRI with an atomic magnetometer

I. Savukov and T. Karaulanov

Los Alamos National Laboratory

Abstract

Ultra-low field (ULF) MRI is a promising method for inexpensive medical imaging with various additional advantages over conventional instruments such as low weight, low power, portability, absence of artifacts from metals, and high contrast. Anatomical ULF MRI has been successfully implemented with SQUIDs, but SQUIDs have the drawback of cryogen requirement. Atomic magnetometers have sensitivity comparable to SQUIDs and can be in principle used for ULF MRI to replace SQUIDs. Unfortunately some problems exist due to the sensitivity of atomic magnetometers to magnetic field and gradients. At low frequency, noise is also substantial and a shielded room is needed for improving sensitivity. In this paper, we show that at 85 kHz, the atomic magnetometer can be used to obtain anatomical images. This is the first demonstration of any use of atomic magnetometers for anatomical MRI. The demonstrated resolution is 1.1x1.4 mm$^2$ in about six minutes of acquisition with SNR of 10. Some applications of the method are discussed. With several measures it is also possible to increase sensitivity to reach resolution 1x1 mm$^2$, which is typical in medical imaging.

Atomic magnetometer, MRI, ultra-low field, anatomical, imaging, low-cost

## I. Introduction

Ultra-low field (ULF) MRI is a promising method of anatomical imaging with possibility to reduce cost, make the system portable, improve contrast [1], enable imaging patients with implants [2], reduce RF power, acoustic noise, and realize some other advantages. Multi-channel detection systems based on low-$T_c$ superconducting quantum interference devices (SQUIDs) have been used to improve sensitivity at very low frequencies of a few kHz together with the pulsed-prepolarization method, and anatomical imaging of the hand [3] and the human brain [4] has been demonstrated. Unfortunately, low-$T_c$ SQUIDs require liquid helium, which is a serious drawback. An alternative method exists based on atomic magnetometers (AMs), which recently achieved high sensitivity at 0.2 fT/Hz$^{1/2}$ level [5], comparable to that of SQUIDs. Previously, AMs have been applied to NMR [6,7] and MRI [8,9] detection in the ULF regime; however, no anatomical imaging has been demonstrated. One problem is low practical sensitivity of the systems used, partially due to external noise at low frequencies.

Because it is difficult to shield magnetic field noise at a low frequency, the detection of MRI signal at higher frequencies, 60-400 kHz, for which most advantages of ULF MRI are still retained, is more attractive. However, the detection of the NMR and MRI signals at such frequencies directly with an AM is problematic [9]. The AM uses the atomic spins as the sensing medium that have a resonance response at frequencies determined by a bias field, 700 kHz/G in the case of the potassium AM. Nuclear spins (protons) precess at Larmor frequencies of 4.26 kHz/G. Thus for detecting an NMR signal at 85 kHz, fields of 20 G and 0.12 G have to be applied separately to the NMR sample and the cell of the AM, respectively. The fields also have to be uniform: the NMR field at the level of 100 ppm and the AM field at 1%. In a previous experiment at 61.5 kHz [7], two required different fields were generated with the aid of a carefully wound long solenoid that contained the NMR sample with the AM positioned outside. The imperfections in winding, even after several attempts in winding with a lathe and after selection of the most field-uniform region, shortened $T_2^*$ relaxation time to ~10 ms. It is too short for anatomical imaging considering practical sensitivity limitations, and an order of magnitude improvement in the field uniformity is needed. However, even more importantly, since the solenoid problem might be in principle solved, a fairly large gradient, 17 mG/cm, has to be applied during the read-out phase as the frequency encoding gradient, which would significantly broaden the AM linewidth to about 12 kHz for a cell of 1 cm size and reduce sensitivity by almost two orders of magnitude.

Alternatively, it is possible to decouple the fields and gradients in the NMR and AM regions with a flux transformer (FT). Ideally, the FT has to be made of superconducting material, but a room-temperature FT can be used as well to avoid cryogens. Recently imaging of a water phantom with a 2-mm resolution at 3.2 kHz has been demonstrated with such a FT [10]. Unfortunately, Johnson noise in the FT negatively affected the detection sensitivity, and it was too low for anatomical imaging. The analysis [10] showed that the noise can be substantially reduced by increasing the detection frequency. In the follow up work, an MRI system operating at 85 kHz has been constructed that achieved much higher sensitivity, and it was used to obtain anatomical images with an atomic magnetometer. This paper presents first anatomical images obtained with an atomic magnetometer.

We demonstrate anatomical images obtained with a single-channel FT+AM system. The demonstrations serve three purposes. The first purpose is to show that anatomical imaging can be realized with an AM in principle. The second purpose is to evaluate the anatomical image quality of the single-channel system for developing applications such as diagnostics of disease of a human hand, or MRI of animals and tissues at ultra-low field to explore high contrast to disease. In some applications, the imaging with the focus on a small area, where anomaly is expected or is suspected, can be needed. The bore size of our system is sufficiently large (16 cm) to implement local diagnostics of a hand, an arm, and animals by moving the area of the interest close to the sensor. For example, the head area of a rat can be imaged for conducting research on diseases of the brain. High contrast at low field can be explored for *in vivo* cancer detection in animals. It is also possible to image a large area, beyond the field of view (FOV) of a small sensor, by translating the object, as we will demonstrate. Unfortunately, this procedure can take a long time, which is proportional to the number of positions. This brings the motivation for building a multi-channel system, with which in 5-10 minutes an image with 1-mm resolution can be obtained, as we will show. The advantages of the multi-channel system have been described previously in Ref. [10]. Thus, the third purpose of the demonstrations is to evaluate the resolution and SNR of a projected multi-channel system, which we plan to construct by combining several one-channel systems.

In addition to the imaging demonstrations, we also characterize the atomic magnetometer in terms of its sensitivity and bandwidth and investigate the properties of the combined FT+AM system. This information can be needed for applications in sensitive detection of rf fields in general, for MRI applications in particular, and for understanding strategies for improvement of sensitivity.

## II. Method

Previously MRI of a water phantom was obtained using an AM+FT sensor at the frequency of 3.2 kHz [10]. The experiment was conducted inside a 2-layer mu-metal shielded room. The technique is applicable to anatomical imaging, but higher frequency of operation is needed to have sufficient SNR and to remove requirement for a shielded room. For this purpose, we constructed an MRI system that operates at 85 kHz without a shielded room. It is based on the idea of pulsed-prepolarizing (Bp) field to enhance the signal detected at ULF frequency, and its various components were described previously in [11]. The MRI system consists of a measurement field (Bm) coil, a set of three gradient coils, the prepolarization Bp coil, an FT+AM detection system, current sources, gradient and rf excitation amplifiers, computer interface boards, and labview software. The Bm coil serves to produce uniform field for imaging. The gradient coils are used to generate three orthogonal gradients of the field component along the Bm direction for 3D imaging and shimming external gradients and imperfections of the Bm coil. The Bp coil, actually a set of four parallel identical coils, is driven by a four-channel power supply with switches to generate a 2kG prepolarization field, which is switched on during the prepolarization cycle and switched off during the MRI sequence. The Bm and gradient coils are identical to those used in Ref. [11]. The FT-AM detection system serves to detect the proton signal. Its principles are described in [10]. A long cylindrical aluminum shell is inserted between the Bp coils and the FT input coil to shield the noise from the Bp coils and ambient noise in the laboratory. Some measures were taken to reduce the effect of transients. The NMR sequences and imaging protocols are similar to those used previously [11]. The gradient strength was increased to improve resolution.

### *Design and test of the AM-FT detection system*

An AM+FT detection system is used to detect the NMR signal. It is constructed to operate at 85 kHz with the sensitivity close to the fundamental limit of the FT. The system is schematically illustrated in Fig. 1. Its AM part consists of an *Atomic cell,* a *Pump beam*, and a *Probe beam*. The circularly polarized *Pump beam* is used to orient atomic spins along its direction, and the linearly polarized *Probe beam* is used to read-out the atomic spin states. The *Pump* and *Probe* beams intersect inside the *Atomic cell* at 90 degrees, defining the active volume of the AM, which is about 1 cm$^3$. The beams are prepared with an optical setup that consists of mirrors, lenses, a polarizer, a polarizing beam splitter, and a λ/4 waveplate. The opticis and lasers producing the *Pump* and *Probe* beams are mounted on an aluminum breadboard. The optical setup is quite similar to that used in [10]. The principles of operation of the AM are explained in [12]. Atomic spins in the cell interact with magnetic field and rotate the *Probe beam* polarization, which is detected with high sensitivity, close to the quantum fluctuation limit, with a polarizing beam splitter and two photodiodes (not shown).

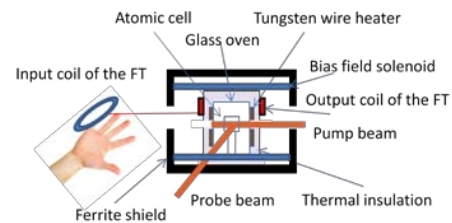

Fig. 1. The AM+FT detection system.

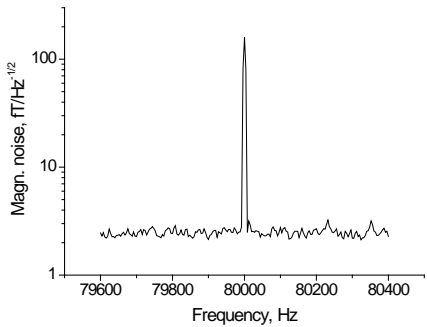

Fig. 2. Atomic magnetometer sensitivity near 80 kHz; the large peak is the calibration signal.

Similarly to nuclear spins, the atomic spins exhibit a resonance response to the oscillating magnetic field. The position of the resonance, which determines the maximum sensitivity of the AM, is adjusted to the NMR frequency with a bias magnetic field.

The AM is enclosed into a cylindrical ferrite shield to reduce the effect of the Earth field, Bm field, gradients, and magnetic noise. The bias field for the AM operation near 80 kHz is generated by the *Bias field solenoid* connected to a DC power supply. Because of the boundary conditions at the ferrite surface, this finite-length solenoid can be approximated as an infinite solenoid. The solenoid performance was acceptable, with the overall bandwith of the atomic magnetometer (FWHM) of 986 Hz, about hundred times smaller than the operation frequency, suggesting the uniformity of better than 1%. The FT consists of two coils: the *Input coil* located outside the *Ferrite shield* to receive the NMR signal from a hand or other imaging objects and the *Output coil* located inside the shield to convert the signal into the magnetic field, which is detected by the AM. The *Output coil* is attached to the external "cold" surface of the AM oven. Home-made *Tungsten wire heaters* are used to bring the temperature of the *Atomic cell* in the oven to 180-200 °C. A low-noise DC power supply is used to provide the current to the heaters, and it is disconnected with an electronic switch during measurements, while reconnected during NMR polarization cycle. The "hot" surface of the oven is surrounded by microporous soft thermal insulation to reduce heat losses and avoid the heating of the ferrite shield.

To minimize noise and achieve long-term stability of the system, we used a distributed-feedback (DFB) laser, *High Powered DFB Single Mode Diode Laser System* from *Sacher Lasertechnik*, as the source of the *Probe beam*. The laser diode was cooled to –12 °C for the operation at 770.8 nm. In the vicinity of 80 kHz, the probe laser noise was found to be close to the photon shot noise. The *Pump beam* was generated by a *Tunable Littrow External Cavity Diode Laser,* also from *Sacher Lasertechnik*. The sensitivity of the magnetometer was found at the level of 2 fT/Hz$^{1/2}$ (Fig. 2). The noise spectrum was also flat suggesting the absence of external rf noise. Careful grounding of the system was needed to exclude substantial external noise. The sensitivity of the FT+AM detection system measured at the *Input coil* was found to be close to the fundamental limit of the FT, determined by Johnson noise, about 1 fT/Hz$^{1/2}$.

## *Flux transformer*

The flux transformer was designed to maximize the sensitivity of the system within constraints of the external oven dimensions. Its *Output coil* was attached to the external surface of the insulated oven. The coil had fixed geometry and the number of turns. The *Input coil* was adjusted in several steps for the best performance in hand MRI experiments. Johnson noise in the *Output coil* (or AC resistance) did not much exceed Johnson noise (AC resistance) of the *Input coil*. This was achieved by adjusting the number of turns of the *Input coil*. The total AC resistance of the FT is

$$R_{FT} = \frac{\omega L_{InputCoil}}{Q_{InputCoil}} + \frac{\omega L_{OutputCoil}}{Q_{OutputCoil}}$$

, where $L_{InputCoil/OutputCoil}$ are the inductances and $Q_{InputCoil/OutputCoil}$ are the quality factors of the Input and Output coils.

For the FT coils used in the hand imaging, $L_{OutputCoil} = 344 \mu H$, $Q_{OutputCoil} = 382$, $L_{InputCoil} = 493 \mu H$, and $Q_{InputCoil} = 140$ giving the total AC resistance 2.36 Ohm. This corresponds to a 12% increase in the Jonson noise due to the *Output coil*, which is almost negligible.

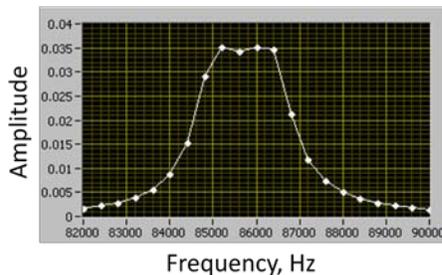

Fig. 3. The frequency response of the AM+FT detection system near 85 kHz used in hand MRI.

In addition to the solution of the field-separation problem, another practical advantage of the FT+AM method, compared to the direct detection method with the AM, is that the rf excitation coil can be made perpendicular to the FT *Input coil* to minimize transients from rf excitations and noise from the excitation coil. We found that the transients were not long, lasting about 4 ms, allowing us to acquire the signal with the AM without a transient-suppression circuit. The noise from the rf excitation coil was insignificant.

Another practical advantage of the FT+AM system is that the bandwidth (BW) can be considerably expanded compared to that of a single coil resonated with a capacitor by shifting magnetometer's peak sensitivity from the FT transfer peak. This can allow use of coils with high Q factors to improve sensitivity without penalty of strong variation of signal sensitivity across the MRI spectrum. For example, the FT *Input coil* used in hand imaging had a Q of 140 at 85 kHz, implying 0.7-level BW of 0.6 kHz, while the BW of the FT+AM system shown in Fig. 3 is about 2 kHz, more than 3 times larger. The FT+AM frequency response is also much more uniform. This allows us to apply stronger frequency encoding gradients to increase resolution. For example, we applied a gradient of 102 Hz/cm (for protons, with 4.26 kHz/G) to achieve the resolution of 1.1 mm, and this would require a 1-kHz bandwidth for a 5-cm image. The BW restriction becomes progressively more important for larger FOV. The sensitivity limited by Johnson noise improves with a Q factor, if all other parameters are fixed, as the square root, so some sensitivity enhancement can be achieved by using very high-Q coils.

Unfortunately, the sensitivity to the NMR signal cannot be indefinitely improved by increasing the number of turns in the *Input coil* for several reasons. First, the size of the coil grows leading to the increase in the effective distance to the imaging object. Second, the external noise becomes the limiting factor, including the Johnson noise from the rf shield needed for reducing external noises, especially generated by the Bp coil. The Johnson noise contribution of the shield was empirically tested as the decrease in a Q factor of the coil when it was inserted into the shield. Another interesting fact about coils is that increasing the number of turns beyond the optimal value reduces their sensitivity due to various parasitic effects such as coil stray capacitance, electrical losses, and the proximity effect. Examples of specific coils are provided later to explore the possibility of sensitivity improvements with frequency.

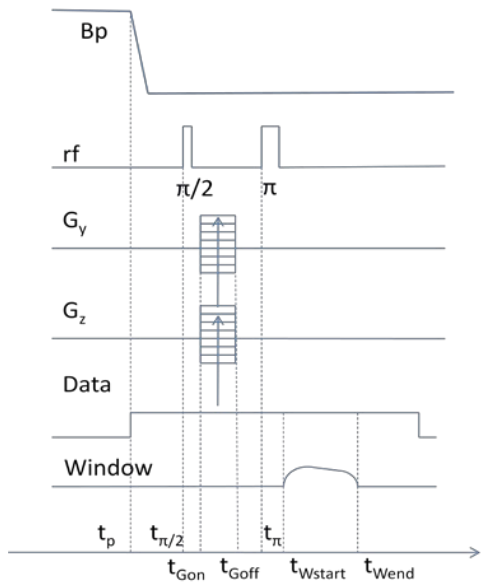

Fig. 4. Pulse sequence: $t_p$=350 ms, $t_{\pi/2}$=$t_p$+49 ms, $t_{Wstart} \approx t_p$ +90 ms. Repetition time is 560 ms. Other parameters are listed in Table

*Transient problem*
To improve SNR as well as resolution and increase the size of imaging objects, we replaced a 1-kG coil, used in our previous experiments [11], with a larger and more powerful 2-kG coil. Unfortunately, such a coil produced much stronger transients. Even though we constructed a switch that could turn off the main current very quickly in 10 ms, slow field drifts persisted delaying the start of the MRI sequence to 120 ms. From the analysis of the field transients, it was determined that the longest transient was produced by the Bm coil, which reacted to switching of the Bp field, because it was oriented in the same direction. After replacing in the Bm circuit the power supply used in [11] with a power amplifier, AE Techtron 7224, set in the current output mode, it became possible to reduce the delay to 69 ms. The amplifier in this mode essentially adjusts voltage to compensate for external and internal influences to maintain constant current in the Bm coil. The BW of the amplifier is sufficiently wide for fast compensation. However, the shorter transients in the Bp coil are not compensated.

In order to further reduce the effects of slow drifts in the strength of uniform magnetic field regardless of their origin, we implemented a down-conversion method with an approximately matching variation in the frequency, parameterized by the amplitude and time constant of exponential decay. The down-conversion method minimized smearing in images and narrowed a Fourier peak of the NMR signal from a small sample. The NMR peak narrowing of a small bottle was observed with delays as short as 20 ms, while smearing removal for a large phantom was effective for delays of 49 ms and longer. When delays between 20 and 49 ms were used, the displacement of the bottle from the center shortened free induction decay (FID) and broadened the spectral peak, indicating the presence of transient gradients. In the future a system for dynamic compensation of gradients can be constructed to reduce the delay time needed for imaging larger objects to gain significantly in SNR for tissues with short relaxation times. Currently, a 49-ms delay has been chosen as optimal.

*NMR frequency and possibility for further sensitivity improvement*

For imaging experiments, the NMR frequency was chosen 85 kHz, as the optimal for the current setup. The limitation comes from the Bm coil, which was not designed for fields greater than 20 G. In general, the sensitivity of the room-temperature FT is expected to improve with the frequency. From the expression for a Q factor and from the Faraday law, for a coil of fixed geometry $SNR \propto \omega/\sqrt{R_{AC}} \propto \omega^{1/2} Q^{1/2}$, where $R_{AC}$ is the AC resistance of the coil at the frequency $\omega$ for which the SNR is calculated. This SNR scaling expression is applicable to coils with different number of turns but of the same diameter. The FT *Input coil* chosen for hand imaging has Q of 140 at 85 kHz. At 200 kHz, a coil of similar size with a slightly smaller number of turns has Q of 146, implying the improvement in SNR 1.6 times. At 400 kHz, this coil has Q decreased to 129, but SNR still grows to 2.1 according to the SNR scaling. On the other hand, at this frequency a similar coil with the number of turns reduced to 50 made of the same wire gives larger Q of 180 and would provide gain in SNR of 2.5 compared to the coil used in the hand imagining experiment at 85 kHz. Thus roughly we estimate that we can improve our current SNR by 1.6 times at 200 kHz and 2.5 times at 400 kHz. The increase in frequency is one direction for the future work, although our Bm coil and its current source need to be improved for field uniformity and stability.

## II. Results and Discussion

We have acquired several MRI scans with the FT+AM sensor to demonstrate the first anatomical images obtained with an atomic magnetometer. The imaging object was chosen the human hand. We used a similar MRI sequence as in Ref. [11], for convenience illustrated in Fig. 4, but with new parameters, which are given in the caption of Fig. 4 and in Table I. There are three sets of imaging parameters. The Fig. 4 caption provides the parameters that are the same in all measurements, while Table I lists the parameters that vary. The first set was used for low-resolution, fast imaging; the second set was used for imaging with intermediate resolution and high SNR; the third set was used for imaging with the highest resolution possible for a reasonable acquisition time and SNR.

Table I. Three sets of imaging and sequence parameters used for demonstrations of anatomical imaging. The parameters that are the same for three sets are given in the caption of Fig. 4. The resolution in horizontal plane was verified with a phantom, but in the depth resolution was calculated using the gradient strength and duration. The resolution slightly changes across the image due to field non-uniformity beyond the first-order gradients.

| set # | $t_\pi - t_{\pi/2}$ ms | $t_{Gon} - t_p$ ms | $t_{Goff} - t_{Gon}$ ms | $G_{ymax}$ Hz/cm | $N_y$ | $G_{zmax}$ Hz/cm | $N_z$ | Gx Hz/cm | $t_{Wend} - t_{Wstart}$ ms | Scan time | Resolution mm$^3$ |
|---|---|---|---|---|---|---|---|---|---|---|---|
| 1 | 25 | 51 | 22 | 116 | 41 | 27 | 7 | 73 | 84 | 2 min 30s | 2x2.5x8.6 |
| 2 | 25 | 51 | 22 | 135 | 51 | 45 | 9 | 88 | 100 | 4 min 12s | 1.4x1.6x5 |
| 3 | 28 | 51 | 25 | 135 | 61 | 45 | 11 | 102 | 100 | 6 min 15s | 1.1x1.4x4.4 |

Imaging with the first set, which gives the fastest speed, is performed to demonstrate a scanning method. This method has some advantages and can be potentially used for medical or research applications. The hand was held in four different positions with a specially designed fixture to acquire four pair-wise overlapping images. These images were combined to obtain a complete image of two fingers of a hand, shown in Fig. 5, to demonstrate imaging possibility of a large object with a sensor that has a limited FOV. The advantage of a small sensor (coil), in the current demonstration of 4 cm average diameter, compared to a large sensor is much higher NMR sensitivity, so that long continuous acquisition is not required to obtain a clear image.

There are two problems with long continuous acquisitions. First, it is difficult for a subject to keep the hand still for a very long time, but if the hand is moved, the image is smeared. Second, our MRI system, and perhaps any system in general, has slow magnetic-field drifts that can cause decrease in the resolution of images and SNR. Apart from this, if the noise is coherent with the NMR signal, averaging method of SNR improvement is not very effective. For example, it can be observed that image quality of a hand in [11] was not much improved by averaging signals of two subsequent scans in time domain, each of duration of 6.7 minutes, contrary to what is expected in the ideal case.

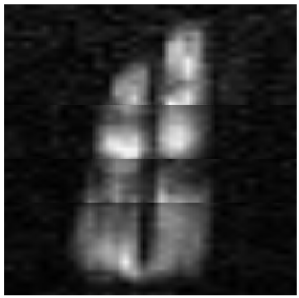

Fig. 5. Composite low-resolution image of two fingers obtained with the first set of parameters.

A small FT *Input coil* is also immune to external noise, including that generated by the rf shield, because the ratio of NMR signal from a voxel to the external noise scales as $1/r_c^3$, where $r_c$ is the radius of the coil. Indeed, we observed that the Q-factors of the coil used in our imaging experiments and the coils of similar size discussed earlier were not much changed when they were inserted into the rf shield, meaning that the Johnson noise in the coils did not increase. In contrast, the Q-factors of larger coils (of radii >10 cm) were quite sensitive to such insertion.

The disadvantage of small sensors (coils) is that they have limited depth sensitivity and FOV. For this reason we chose the *Input coil* of the average radius of 2 cm to have enough depth sensitivity for fingers. The scanning method of course can be used to cover a larger area to remove FOV limitation. The total time required scales with the number of scans, while the quality for the composite image is the same as that of a single scan. If a multi-channel system were implemented with the FT+AMs, then the image of the same quality would be obtained within the time of a single scan. The multi-channel implementation would dramatically reduce imaging time for large objects.

One possible application of the scanning method is to provide a preliminary anatomical image for subsequent zooming in on the area of interest, such as an anomaly. This approach can be of practical value, because very often anomalies are localized, and once detected or suspected, they can be studied with higher resolution using longer scans. Another scenario is that the anomaly location is revealed with traditional imaging methods, high-field MRI or X-ray projection imaging, but the study needs to be continued on the progress of disease, for example, to evaluate response of the disease to treatment. It might be impractical to repeat imaging with high-field MRI machines due to price and limited availability or X-ray machines due to exposure to radiation. The availability can be especially an issue for patients from remote unpopulated areas. Our system being inexpensive, portable, and radiation-free can be a great asset for such a follow up study. The only concern is low image quality. To address this issue, we have investigated the limits on the resolution and SNR when they are separately optimized without resort to a very long scan time. The results of this study are presented next.

In Fig. 6, we show images of 2 adjacent slices of the middle area of the fingers shown in Fig. 5 with fairly high SNR and resolution of 1.4x1.6 mm$^2$ that were acquired in 4 min 12 sec. They are obtained using the second set of parameters listed in Table I. The maximum SNR evaluated for the right image from a one-dimensional intensity profile (Fig. 7) in the horizontal direction is 20. The slice thickness was set to 5 mm, and two adjacent slices show image variation between slices. With thinner slices than in Fig. 5 the contrast has been increased, because images of adjacent slices do not overlap well, and if superimposed would obscure features. Further reduction in the slice thickness would not improve visibility due to reduction in SNR. Higher resolution reveals new features unseen in Fig. 5.

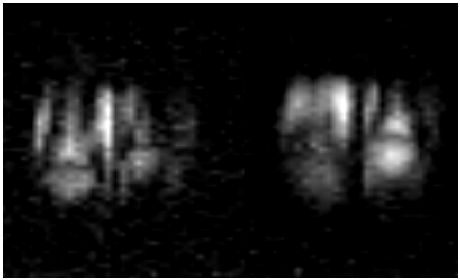

Fig. 6. Images of middle area of fingers shown in Fig. 5; two adjacent slices each 5 mm thick with in-plane resolution 1.4 x 1.6 mm$^2$; acquisition

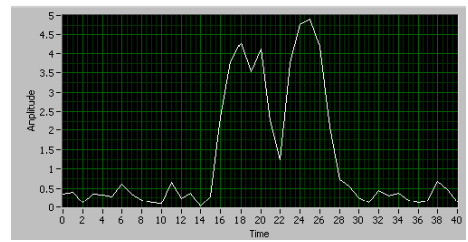

Fig. 7. The one-dimensional profile of the right image in Fig. 6 in the horizontal direction passing the brightest spot. The signal peaks at about 5 units, and noise is about 0.25 units, so the maximum SNR is 20.

Images shown in Fig. 6 have already revealed multiple fine features of anatomy, but the resolution can be still improved within constraints of the minimal acceptable SNR and the scan time. The next images, shown in Fig. 8, are refinement on the images of Fig. 6 and were acquired using the third set of parameters to demonstrate the ultimate performance of our system in resolution, which was determined 1.1x1.4 mm$^2$ in the image plane and 4.5 mm in the slice-selection direction. The left image obtained in a single scan in 6 min 15 sec has some distortions due to noise fluctuations, so we recorded a second image using identical settings and combined it with the first one by adding intensities. The resulting image on the right has improved quality. This is an indication that now SNR is a limiting factor for visibility for shorter scans. It is interesting to note that the images reveal sharp boundaries and show a narrow gap between bones in the joint of the finger.

The demonstrated anatomical images serve at least two purposes. They give an idea of image quality, which can help radiologists to evaluate the medical utility of the ULF MRI method developed here. On the other hand, from the images quantitative information about SNR can be extracted for comparison with other methods and objective evaluation for future improvement. The resolution, which cannot be judged accurately from anatomical images, was evaluated from images of a phantom, which have well-defined geometrical features: multiple round holes located on a regular 2D rectangular grid with 10-mm separation between centers of the holes (Fig. 8). Only the phantom images obtained with the third set of imaging parameters is shown (Fig.10); in other cases the images are quite similar, although of lower resolution, so it is enough to show only one case. In all cases, the phantom images were used to determine the resolution. In Fig. 10, we show both the image with original pixel size determined by FFT (left) and the image interpolated to higher resolution (by a factor of 10). The image with the original pixel size is used for the resolution determination, while interpolated images are used to improve the perception of anatomical features. In the case of the phantom, the round holes are seen as round holes after the interpolation.

The phantom image also shows that FOV is about 5 cm in the horizontal direction and 3 cm in the vertical direction. Because the FT *Input coil* is round, this asymmetry cannot arise from the sensor geometry but rather should be attributed to the finite amplitude of the rf excitation field. This can be solved by increasing the magnetic field of rf pulses, but our hardware has to be modified.

Apart from potential applications in localized diagnostics, as discussed earlier, the illustrated images can be also used to evaluate image quality of a multi-channel system that can be constructed in the future for acceleration of image acquisition. The image of a hand, which does not exceed 10x20 $cm^2$, can be obtained with 8 parallel sensors each 5 cm in diameter with the FOV of 5x5 $cm^2$, assuming that the current limitation of FOV in the vertical direction will be improved. In 6 minutes and 15 seconds we should be able to obtain a complete hand image with the resolution 1.1x1.4 $mm^2$ similar in resolution and SNR to that shown in Fig. 8. This will be substantial improvement in quality over the previous demonstrations [11].

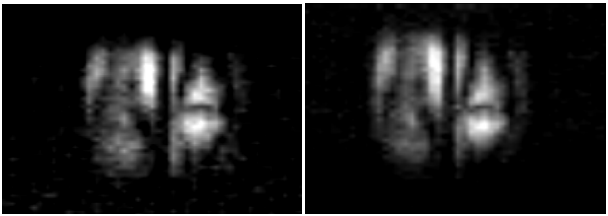

Fig. 8. The anatomical images of the middle area of the fingers shown in Fig. 5 with resolution 1.1x1.4 $mm^2$ (4.4-mm slice) obtained in 6min 15 sec of acquisition, the left image, and in 12 min 30 sec, the right image. The right image is the intensity average of two images obtained in 6 min 15 sec each.

The current demonstrations do not exemplify yet the best possible image quality that can be obtained with the ULF MRI method based on the FT+AM detection system. Two-time longer scans, for example, can increase the SNR by a factor of 1.4, and this is one measure that can be taken for improving the resolution to 1x1 $mm^2$. On the other hand, with hardware modifications, SNR and resolution can be improved by reducing the effect of transients and by increasing the detection frequency. By reducing transients, and we already have done some preliminary feasibility tests, the delay time can be shortened by 29 ms to increase SNR 1.6 times for tissues with $T_1$ of 60 ms. This should be possible by implementing dynamic gradient and field compensation. An additional factor of 2 is expected from raising frequency to 400 kHz, as tests of Q-factors of coils suggest. The improvement with these two last measures in SNR is 3.2. This SNR gain can be used to increase the resolution to a typical in medical practice of 1x1 $mm^2$ and reduce the scan time to 3 minutes. With these improvements, even a single-channel system can be used to obtain the image of a whole hand in 24 minutes with the resolution 1x1 $mm^2$.

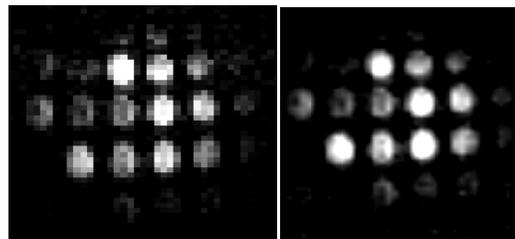

Fig. 9. Phantom images with 1.1x1.4 $mm^2$ resolution. The left image has initial pixel size, but the left image was up sampled in the same way as the anatomical images.

## III. Conclusion

We have demonstrated first anatomical MRI with an atomic magnetometer. We achieved resolution 1.1x1.4 mm in 6 minutes 15 seconds with SNR about 10. The resolution approaches that typical in medical MRI. We anticipate some applications of the current one-channel system in diagnostics of hand diseases based on the two-step procedure described above. First, the low-resolution but fast scanning method can be used to obtain a complete image to search for suspicious features. Then, high-resolution image can be acquired with the focus on the location of the anomaly. Currently, the image of 1.1x1.4 mm$^2$ resolution can be obtained in 6-12 minutes to study this anomaly with FOV of 3x5 cm$^2$. The first step can be also replaced with some other diagnostics, conventional MRI, X-ray imaging, etc., and the follow-up study can be performed with focused imaging. Alternatively, a multi-channel system can be constructed in the future to obtain high-resolution image in a single scan. It will be more complicated, but considering very high price of conventional MRI scanners, this extra complexity will not be a problematic issue. The results obtained here with one-channel system can be used to project the image quality of a multi-channel system, which can have similar FT+AM sensors. The advantage for hand imaging is the acceleration by a factor of 8 if the whole hand image is needed.

We have already made substantial progress in sensitivity and resolution over the previous ULF MRI experiments, but there is still much room for improvement. For example, with some modifications in hardware, effects of transients from the prepolarization coil can be reduced and the frequency of the detection can be increased bringing overall gain in SNR of 3.2. This would enable us to do imaging with 1x1 mm$^2$ resolution just in 3 minutes.

Finally, anatomical imaging is demonstrated with atomic magnetometers for the first time, which is an important milestone for applications of atomic magnetometers in biomedical imaging. The results are quite encouraging and further improvement is expected. New medical applications can be enabled, but it will be necessary to conduct some research on diagnostics of anomalies, emphasizing the potential gain in the contrast in the ULF regime. Various advantages such as low cost, portability, simplicity, safety, open design, high contrast, and absence of susceptibility artifacts can be attractive for novel applications of the system. Apart from applications in human diagnostics, the system can be used for imaging animals (veterinary applications or research) and tissues to diagnose and study diseases. Because of low cost, safety, and small size, the system can be also attractive to for world-wide academic research and teaching.

## Acknowledgement
This work is sponsored by NIH grant 5 R01 EB009355.